\documentclass[12pt]{article}

\usepackage{amsmath,amssymb}
\usepackage{bm}
\usepackage{graphicx}

\makeatletter

\@addtoreset{equation}{section}
\makeatother

\newcommand{\beq}{\begin{equation}}
\newcommand{\eeq}{\end{equation}}

\newcommand{\bC}{\ensuremath{\mathbb{C}}}

\newcommand{\bP}{\ensuremath{\mathbb{P}}}

\newcommand{\bR}{\ensuremath{\mathbb{R}}}

\newcommand{\bZ}{\ensuremath{\mathbb{Z}}}

\newcommand{\scZ}{\ensuremath{\mathcal{Z}}}

\begin{document}

\baselineskip=18pt  % a la harvmac
\baselineskip 0.8cm

\begin{titlepage}

%% Set the number of the title with 0
\setcounter{page}{0}

% change the footnote symbol
\renewcommand{\thefootnote}{\fnsymbol{footnote}}

%------------------
\begin{flushright}
{\tt CALT-68-2746\\
IPMU09-0091\\
UT-09-18\\}
\end{flushright}
%---------------------

\vskip 1cm

\begin{center}
{\LARGE \bf
 Wall Crossing and M-Theory}

\vskip 2cm

{
%\authors
{\large Mina Aganagic$^1$, Hirosi Ooguri$^{2,3}$, Cumrun Vafa$^4$
%\\
%\medskip
and Masahito Yamazaki$^{2,3,5}$}
}

\vskip 1cm

{
\it
%\large
%affiliations
$^1$Center for Theoretical Physics, University of California, Berkeley, CA 94720, USA\\
\vskip 0.05cm
$^2$California Institute of Technology, Pasadena, CA 91125, USA\\
\vskip 0.05cm
$^3$ IPMU, University of Tokyo, Chiba 277-8586, Japan\\
\vskip 0.05cm
$^4$Jefferson Physical Laboratory, Harvard University,  Cambridge, MA 02138, USA\\
\vskip 0.05cm
$^5$Department of Physics, University of Tokyo, Tokyo 113-0033, Japan\\
}

\end{center}

\vspace{1.5cm}

%-----------------------------------------
\centerline{{\bf Abstract}}
\medskip
\noindent
We study BPS bound states of D0 and D2 branes on  
a single D6 brane wrapping a Calabi-Yau 3-fold $X$. 
When $X$ has no compact 4-cyles, the 
BPS bound states are organized into a free field Fock space, whose 
generators correspond to BPS states of spinning M2 branes in M-theory 
compactified down to 5 dimensions by a Calabi-Yau 3-fold $X$. 
The generating function of the D-brane bound states is 
expressed as a reduction of the square of the topological string partition function,
in all chambers of the K\"ahler moduli space. 

\end{titlepage}
\setcounter{page}{1} % don't number title page

%%%%%%%%%%%%%%%%%%%%%%%%%%%%%%%%%%%%%%%%%%%%%%%%%%%%%%%%%%%%%%%%%%%%%%

\section{Introduction}

The topological string theory gives solutions to a variety of 
counting problems in string theory and M-theory.
From the worldsheet perspective, the A-model topological string partition
function  ${\cal Z}_{\rm top}$ generates the Gromov-Witten invariants, which
count holomorphic curves in a Calabi-Yau (CY) 3-fold $X$. On the other hand,
from the target space perspective, ${\cal Z}_{\rm top}$ computes
the Gopakumar-Vafa (GV) invariants, which count BPS states of
spinning black holes in 5 dimensions
constructed from M2 branes in M-theory on $X$ \cite{GV1}.
Moreover, the absolute-value-squared $|{\cal Z}_{\rm top}|^2$ has been
related to the partition function
of BPS black holes in 4 dimensions, which are bound states of D branes
in type II string theory
on $X$ \cite{OSV}.

The topological string partition function ${\cal Z}_{\rm top}$ also
counts the numbers of D0 and D2
brane bound states on a single D6 brane on $X$, namely the
Donaldson-Thomas (DT) invariants defined
in \cite{DT,Thomas}. The relation between the GV invariants
 and DT invariants was suggested and
formulated in \cite{INOV,MNOP}, and its physical explanation was given in \cite{DVV} using the 4D/5D
connection \cite{GSY}. More recently, a mathematical proof of the GV/DT correspondence was given in
\cite{MOOP} when $X$ is a toric CY 3-fold.

However, the number of BPS states has background dependence. 
As we vary moduli of
the background geometry and cross a wall of marginal stability, 
the number can jump
\cite{GMN,KS}.
In this paper we will generalize the results of \cite{DVV} to include 
the background dependence of
the M-theory computation. 
We show BPS bound states are organized into a free field Fock space, whose 
generators correspond to BPS states of spinning M2 branes in M-theory 
compactified down to 5 dimensions by a Calabi-Yau 3-fold $X$. This
enables us to write the generating function 
${\cal Z}_{\rm BPS}$ of BPS bound states of D-branes
as a reduction of the square of the topological string partition function,
\beq
   {\cal Z}_{\rm BPS} = {\cal Z}_{\rm top}^2 |_{\rm chamber}, \label{theformula}
\eeq
in an appropriate sense described in the following, 
in all chambers of the K\"ahler moduli space. 
Our results apply to the BPS counting
for an arbitrary CY (whether toric or non-toric) without compact 4-cycles. 

For the conifold, the change
of the numbers of BPS states across a wall of marginal stability
has been studied by physicists in \cite{JM,CJ} (see also \cite{DG}) and
mathematicians \cite{Young1,NN}. The case of generalized conifold geometries was
studied in \cite{Nagao}. 
The formula \eqref{theformula} derived 
from the perspective of M-theory reproduces these results. 
Our results also provide a simple derivation of the ``semi-primitive" wall
crossing formula of Denef and Moore \cite{DM}, in the present context.

The rest of the paper is organized as follows. In section 2, we will explain the basic idea
to use M-theory to count bound states of a single D6 brane with D0 and D2 branes on a CY
3-fold. In section 3, we will describe the counting procedure in more detail and derive
the generating function for the numbers of BPS bound states using a free field Fock space
in any chamber of the background K\"ahler moduli space. 
In section 4, we will compare the Fock space picture with
the known results for the the resolved
conifold and its generalizations. Finally, in section 5 we point out that
our results give a derivation of the Denef-Moore ``semi-primitive'' wall-crossing formula in the present context.

\section{The basic idea}

In this section we will explain the basic idea.  We will apply this
idea, in the following sections, to find a concrete expression for
BPS state degeneracies in various chambers for CY 3-folds with no
compact 4-cycles.

We are interested in counting the BPS partition function of one D6 brane
bound to arbitrary number of D2 and D0 branes.  The idea is
the following:  In M-theory, the D6 brane lifts to the Taub-NUT space with
the unit charge.
D2 branes are M2 branes transverse to the $S^1$, and D0 branes are
gravitons with Kaluza-Klein momenta along the $S^1$.
The Taub-NUT space is an $S^1$ fibration over $\bR^3$,
and $S^1$ shrinks at the position of D6. Thus the problem of
finding bound states to the D6 brane becomes simply the problem of finding BPS
states in the Taub-NUT geometry.  Suppose we have BPS states for flat
$\bR^{4,1}$ background.  Then for each such BPS state we can consider the
corresponding possible BPS states in the Taub-NUT geometry.  For each
single particle BPS state we can consider its normalized wave functions
in this geometry.  Such states would constitute BPS states which in the type IIA
reduction correspond to BPS particles bound to the D6 brane.  However,
this would only constitute single particle BPS states bound to the D6 brane.

Now consider multiple such particles in the Taub-NUT background.  This problem
may sound formidable, because now we will have to consider the interaction
of such particles with each other and even their potentially forming
new bound states.  We will now make the following two assumptions:

\medskip

\noindent
{\bf Assumption 1}:  We can choose the background moduli of CY as well
as the chemical potential so that a maximal set of
BPS states have parallel central charge and thus exert
no force on one another.  Therefore, at far away separation, the bound states
correspond to single particle wave function in the Taub-NUT geometry.

\noindent
{\bf Assumption 2}:  The only BPS states in 5D are particles.  In other words
there are no compact 4-cycles in the CY and thus we can ignore BPS {\it string
states} obtained by wrapping 5 branes around 4-cycles.

\medskip

Assumption 1 can be satisfied as follows:  Consider the Euclidean geometry
of M-theory in the form of Taub-NUT times $S^1$, where we have compactified
the Euclidean time on the circle.  The BPS central charge for M2 branes wrapping
2-cycles of CY, but with no excitation along the Taub-NUT, is given by
$$Z(M2)=iA(M2)- C(M2),$$
where $A(M2)$ denotes the area of the M2 brane and $C(M2)$ corresponds to the
coupling of the M2 brane to the 3-form potential turned on along the CY 2-cycles
as well as the $S^1$ of the Taub-NUT. However we need to include excitations
along the Taub-NUT.  As discussed in \cite{DVV} these are given by the momenta
along the Taub-NUT circle.  Let us denote the total momentum along the circle
by $n$ (as we will review in section 3, this can arise both due to internal spin
as well as the orbital spin in the $SU(2)_L\subset SO(4)_{\rm rotation}$). Let us denote the radius of the Taub-NUT
circle by $1/R$.  In this case the central charges
of the BPS M2 brane becomes\footnote{
Note that $C(M2)$ is periodic with period $1/R$. To see this, note that we can view it as a holonomy of the gauge field obtained by reducing the 3-form on the 2-cycle of CY, around the Taub-NUT circle. The holonomy of a gauge field on a circle of radius $R$ is periodic, with period $1/R$. In terms of the IIA quantities, we have
$$
C(M2) = B(D2)/R,
$$
where $B(D2)$ is the NS-NS B-field through the 2-cycle in IIA on the CY wrapped by the corresponding D2 brane (which has periodicity $B\to B+1$). We are denoting by $1/R$ the central charge of the D0 brane. As such, it does not have to be positive, and in fact it does not have to be real either. A better way to think about this is that the quantity we denote by $1/R$ is proportional, up to a complex constant, to the inverse radius of the M-theory circle. For simplicity of the notation, we will identify the two, but this fact has to be kept in mind,
Note also that relative to \cite{JM} we are keeping the D6 brane charge fixed, and varying the D0
and D2 brane central charges.}
$$Z(M2,n)=iA(M2)-C(M2)-n/R.$$

To satisfy assumption 1, we need to make sure that differently wrapped
M2 branes all have the same phase for $Z$.  This in particular means that
we need to choose the K\"ahler classes so that the 2-cycles of CY have
all shrunk to zero size, $i.e.$
$A(M2)=0$ for all the classes.  Even though this may
sound singular and it could lead to many massless states, by turning
on the $C(M2)$ we can avoid generating massless states in the limit.
The condition that different states have the same central charge
is simply that
\beq
C(M2)+n/R>0.
\eeq
Note that, in going to type IIA, this condition is simply the statement
that the $B$ fields are turned on along 2-cycles of CY and the M2 branes
wrapping them will have $B(D2)$ 0 branes induced.  Moreover $n$, being
the momentum along the Taub-NUT translates to D0 brane charge and as long
as the net number of 0 branes is positive, they correspond to BPS states
of the same type, $i.e.$ preserving the same supersymmetry.

%It will be
%convenient below to write $exp(-C(M2))=Q^\beta$ where $\beta$ denotes the class
%of M2 brane inside the CY and $Q=exp(-C)$ (where we are using a shorthand
%notation suppressing indices for $H^2$ classes).  It is also convenient
%to denote $q=exp(-1/R)$.  Then such a BPS state
%contributes to the partition function by the factor
%
%$$Q^\beta q^m$$
%
%and we have the positivity condition noted above for mutually
%BPS condition:
%$$Q^\beta q^m<1.$$

Now we are ready to put together
all these mutually BPS states as a gas of particles
in the Taub-NUT geometry.  By the fact that they are mutually BPS, they
will exert no force on one another.  Moreover, as long as they are far away,
we can simply consider the product of the individual wave functions.
One may worry what happens if they come close together.  Indeed
they can form bound states, but that is already accounted for by
including all single particle bound states of M2 brane.  Here
is where the assumption 2 becomes important:  If we in addition had
4-cycles, then wrapped M5 branes along 4-cycles, which also wrap the
$S^1$ of Taub-NUT can now form new bound state with the gas of M2
brane particles on the Taub-NUT.  But in the absence of 4-cycles
of CY, we can simply take the single particle wave functions
(taking their statistics into account) and write the total degeneracy
of such BPS states, by taking suitable bosonic/fermionic creation
operators, one for each state satisfying $C(M2)+n/R>0$. Finally, while the assumption 1 is satisfied only for special backgrounds where $A(M2)$ vanishes, the degeneracies are guaranteed to be the same everywhere within a given chamber, and independent of this choice. This is
all we need to compute all the degeneracies of BPS states in various chambers
as we will show in the following sections.

\section{BPS state counting and wall crossing}

We will use this section to spell out, in a little more detail,
how to use M-theory to compute the degeneracies of
one D6 brane on $X$ bound to D2 branes wrapping 2-cycles in $X$ and D0 branes.
The D6-D2-D0 partition function is the Witten index\footnote{
Here we are ignoring the fermionic zero modes in the 4 non-compact directions. Otherwise, additional factors need to be inserted to absorb these.}

$$
{\rm Tr}[(-1)^F e^{-\epsilon H}]
$$
of the theory on
$$
X \times \bR^3 \times S_t^1,
$$
where we have compactified the Euclidian time on a circle of radius 
$\epsilon$. The type IIA geometry with one D6 brane lifts to to M-theory on
$$
X \times {\rm Taub} \hbox{-} {\rm NUT} \times S_t^1,
$$
where the asymptotic radius of the Taub-NUT circle $R$ is related to IIA string coupling.
Since the D6 brane is geometrized,
the computation of the BPS bound states of D2 branes and D0 branes with D6 brane
lifts to a question of computing the degeneracies of
$M2$ branes with momentum around the Taub-NUT circle.

Suppose we know the degeneracies of M-theory in the
$$
X \times \bR^4 \times S^1.
$$
This corresponds to taking the
$R\to \infty$
limit, where the Taub-NUT just becomes $\bR^4$.
As is clear from the previous section, at {\it fixed} $B$ the degeneracies are {\it unchanged} by varying
$R$ since no states decay in the process --- all the central charges simply get re-scaled.
Thus, the knowledge of these allows us to compute the degeneracies on
$X \times$Taub-NUT$\times S^1$ background as well.

The Kaluza-Klein momentum around the Taub-NUT circle gets identified, in terms of the theory in the $R\rightarrow \infty $ limit, with the total spin of the M2 brane. This can be understood by comparing the isometries of the finite and the infinite $R$
theory, as explained in \cite{DVV}.
We can view taking $R$ to infinity as zooming in to the origin of the Taub-NUT. The isometry group is the rotation group $SO(4) = SU(2)_L \times SU(2)_R$ about the origin of $\bR^4$. The $SU(2)_R$ is identified with the $SO(3)$ that rotates the sphere at infinity of the $\bR^3$ base of the Taub-NUT. Moreover, the rotations around the $S^1$ of the Taub-NUT, end up identified with the
$$
U(1) \subset SU(2)_L.
$$
Thus, the Kaluza-Klein momentum, is identified with the total $J^{L}_z$ spin of the M2 brane on $\bR^4$.

Now, let
$$N_{\beta} ^{(m_L, m_R)}$$
be the degeneracy of the {\it 5-dimensional BPS} states of M2 branes of charge $\beta$ and spin the intrinsic $(2j^z_L, 2 j^z_R) =(m_L, m_R )$
(where the spin refers to the spin of the highest state of the multiplet).
To get an index, we will be tracing over the $SU(2)_R$
quantum numbers, so we get a net number
$$
N_{\beta}^{m_L} = \sum_{m_R} (-1)^{m_R} N_{\beta}^{(m_L, m_R)}
$$
of 5D BPS states, of the fixed $SU(2)_L$ spin $m_L$.

Each such 5D BPS particle can in addition have excitations on $\bR^4$. Namely, for each 5D particle we get a field
$$\Phi(z_1,z_2)$$
on $\bR^4$ with $z_{1,2}$ as the complex coordinates. In the usual way, the modes of this field
$$
\Phi(z_1,z_2) = \sum_{\ell_1, \ell_2} \alpha_{\ell_1,\ell_2} z_1^{\ell_1} z_2^{\ell_2}
$$
correspond to the ground-state wave functions of the particle with different momenta on $\bR^4$. (We are suppressing a Gaussian factor that ensures the wave functions are normalizable).
Since $U(1)\in SU(2)_L$ acts on $z_1,z_2$ with charge $1$, the particle corresponding to
$$
{\alpha}_{\ell_1,\ell_2}
$$
carries, in addition to the M2 brane charge $\beta$ and intrinsic momentum $m$, a total angular momentum,\footnote{In the present case, we are restricting to CY manifolds with no compact 4-cycles. When the CY is furthermore toric, as in the cases discussed in sections \ref{sec.conifold} and \ref{sec.gc}, the genus of the target space curve wrapped by the M2 branes vanishes. This means that the intrinsic spin of all the M2 branes vanishes as well.}
%In that context, the intrinsic spin of all the M2 branes vanishes. The intrinsic spin is correlated with the genus of the target space curve wrapped by the M2 branes. When this vanishes, as is always the case for toric CY with no compact 4-cycles, the intrinsic angular momentum vanishes as well.}
%
$$n = \ell_1+\ell_2+m.
$$

Which of these 1-particle states are mutually BPS? The answer depends on the background, and a priori, we need to consider particles in four dimensions coming from both the M2 branes and the anti-M2 branes in M-theory.
Along the slice in the moduli space we have been considering, the central charges of the particle with M2 brane charge $\beta$ and total spin $n$ is
$$
Z(\beta, n) = \beta C + n/R =  (\beta B + n)/R.
$$
The states with
\beq\label{cc}
Z(\beta,n)>0
\eeq
all preserve the same supersymmetry and bind to the D6 brane (we could have picked the opposite sign, and than the particles would bind to anti-D6 branes).
For example, for 
$$
B>0, \qquad R>0
$$
along side M2 branes with $\beta>0$, and for sufficiently large $n$ also the anti-M2 branes with $\beta<0$ have positive $Z>0$ and contribute to the BPS partition function. So in general we need to consider both signs of $\beta$. It is important to note that the degeneracies $N_{\beta}^{m}$ of the 5D particles are independent of the background. The choice of background only affects which half of the supersymmetry the states preserve.

Now, we can put these all together and compute the BPS partition function in a given chamber. Simply, in each chamber, the BPS partition function is the character in the Fock space of single particle states
preserving the same supersymmetry! In fact, a useful way to go about computing the partition function in steps:
\vskip 0.3cm
\noindent
{\bf Step 1.\;}  Start with the unrestricted partition function -- the character
$$
{\cal Z}_{\rm Fock} = {\rm Tr}_{\rm Fock} \; q^{Q_0} Q^{Q_2}
$$
in the full Fock space.
The oscillators of charge $\beta$ and intrinsic spin $m$ and arbitrary 4d momenta contribute a factor
\beq\label{factor}
\prod_{\ell_1+\ell_2=n} (1- q^{\ell_1+\ell_2 +m } Q^{\beta})^{N_{\beta}^m}
=
(1-q^{n+m} Q^{\beta})^{n N_{\beta}^m}.
\eeq
In addition, both the M2 branes, and the anti-M2 branes contribute, and the total character is
$$
{\cal Z}_{\rm Fock} = \prod_{\beta,m} \prod_{n=1}^{\infty}\,
(1-q^{n+m} Q^{\beta})^{n N_\beta^{m}}
$$
\vskip 0.3cm

\noindent
{\bf Step 2.\;} The 5d degeneracies $N_{\beta}^m$ of M-theory on $X\times R^{4,1}$ are computed by the topological string partition function on $X$ \cite{GV1, GV2} .  This allows us to write %
$$
{\cal Z}_{\rm Fock} = {\cal Z}_{\rm top}(q,Q) {\cal Z}_{\rm top}(q,Q^{-1}).
$$
In particular, the knowledge of topological string amplitude allows us to compute the BPS degeneracies in any chamber.

The topological string partition function has an expansion
$$
{\cal Z}_{\rm top}(q,Q) = M(q)^{\chi(X)/2} \prod_{\beta>0, m} \prod_{n=1}^{\infty}
(1-q^{m+n} Q^{\beta})^{n N_m^{\beta}}.
$$
where $q$ and $Q$ are determined by the string coupling constant $g_s$
and the K\"ahler moduli $t$ by $q=e^{-g_s}$ and
$Q = e^{-t}$. The MacMahon function $M(q)$ is defined by
$$M(q)= \prod_{n=1}^{\infty}(1-q^n)^{-n}.$$ 
Above, $\chi(X)$ is the Euler characteristic of $X$. Note that topological string involves only the M2 states with positive $\beta>0$. On the other hand, the full Fock space includes also anti-M2 branes. Since M2 branes and anti-M2 branes are CPT conjugates in 5d, this gives another factor of ${\cal Z}_{\rm top}$ with $Q \rightarrow Q^{-1}$. 

Note that we also have states with $\beta=0$. 
These are the pure KK modes, the particles with no M2 brane charge. 
To count the number of BPS states of this type,
we note that, for each $R^4$ momentum $(l_1,l_2)$ we get a classical particle 
whose moduli space is the Calabi-Yau $X$. Quantizing this, we get a particle 
for each element of the cohomology of $X$. On a $(p,q)$ form $SU(2)_R$ acts 
with the Lefshetz action, and $SU(2)_L$ acts trivially. We get that $m_R$ 
eigenvalue of a $(p,q)$ form on $X$ is $m_R = p+q - 3$. Therefore,
the pure KK modes contribute with $N_{\beta=0} = -\chi(X)$. This agrees
with the power of the MacMahon function $M(q)$ we get from 
${\cal Z}_{\rm top}(q,Q) {\cal Z}_{\rm top}(q,Q^{-1})$.

\vskip 0.3cm

\noindent
{\bf Step 3.} We identify the walls of marginal stability as places where,
the central charge vanishes 
for one of the oscillators contributing to ${\cal Z}_{\rm top}(q,Q)$ or  
${\cal Z}_{\rm top}(q,Q^{-1})$.

\vskip 0.3cm
\noindent
{\bf Step 4.\;} In any chamber, the BPS partition function is a restriction 
of ${\cal Z}_{\rm Fock}$ to the subspace 
of states that satisfy $Z(\beta,n)>0$ in that chamber.
\begin{align}
{\cal Z}_{\rm BPS}({\rm chamber}) =& {\cal Z}_{\rm Fock}|_{\rm chamber}, \\
 = & {\cal Z}_{\rm top}(q,Q) {\cal Z}_{\rm top}(q,Q^{-1})|_{\rm chamber} \label{ZChamber}
\end{align}

There is a simple way to keep track of the chamber dependence. For the book-keeping purposes, it is useful to identify the central charge with the chemical potentials. Then, in a given chamber, the BPS states are those for which
$$
q^n Q^{\beta}<1
$$
where $n=m+k$ is the total spin. As we vary the background, and cross into a chamber where this is no longer satisfied for some $(n,\beta)$ in ${\cal Z}_{\rm top}(q,Q)$ or in ${\cal Z}_{\rm top}(q, Q^{-1})$, we drop the contribution of the corresponding oscillator. 
\vskip 0.3cm

For example, consider some special cases. When
\beq 
R>0, \qquad B \rightarrow \infty, \label{DTChamber}
\eeq
for all K\"ahler classes, $Z(\beta,n)= (\beta B +n)/R>0$ implies that
$$
\beta>0.
$$
In this case, only M2 branes contribute to the partition function.
This is the chamber discussed in \cite{DVV}. By taking the limit
\eqref{DTChamber} in \eqref{ZChamber}, we find
$$
{\cal Z}_{\rm BPS}(R>0, B \rightarrow \infty)  = 
{\cal Z}_{\rm DT}(q,Q) = M(q)^{\chi/2} {\cal Z}_{\rm top}(q,Q).
$$
The partition function in this chamber computes DT invariants.
In \cite{MOOP}, it was shown that, for a toric CY,
${\cal Z}_{\rm BPS}$ is equal to ${\cal Z}_{\rm top}$ up to a factor
which depends only on $q$. 
Here we derived the relation between ${\cal Z}_{\rm BPS}$ and
${\cal Z}_{\rm top}$ including the factor of $M(q)^{\chi/2}$. 

On the other hand, when $0 < B \ll 1$, the BPS partition function 
is given by 
\beq
   {\cal Z}_{\rm BPS}(q,Q) =
{\cal Z}_{\rm NCDT}(q,Q) =  {\cal Z}_{\rm top}(q,Q)
{\cal Z}_{\rm top}(q,Q^{-1}). \label{NCDTgeneral}
\eeq
This gives the non-commutative DT invariants studied in \cite{Szendroi,MR,OY1}. When
$X$ is toric, the partition function is computed using the crystal
melting picture \cite{MR,OY1}, generalizing the previous result of \cite{ORV,INOV} for $\bC^3$. In \cite{OY2}, it was shown that the
thermodynamic limit of the partition function of the crystal
melting model gives the genus-$0$ topologica
string partition function. This result was mysterious since the relation
between ${\cal Z}_{\rm top}$ and ${\cal Z}_{\rm BPS}$ was supposed to hold
in the DT chamber discussed in the previous paragraph. 
We now understand why there is such a relation in the non-commutative
DT chamber also as in \eqref{NCDTgeneral}.

%Now let us consider the case when $B$ takes a finite positive value. 
%Let us take $B_*>0$, and consider the chamber, 
%
%$$R>0, \qquad B_*<B<B_*+1.
%$$
%
%In this case, M2 branes with arbitrary $n$ contribute, so we get a 
%full factor of ${\cal Z}_{\rm top}(q,Q)$. In addition, 
%anti-M2 branes for sufficiently large $n$ can contribute as well. 
%We can then  write,
%
%$$
%{\cal Z}_{\rm BPS}(R>0, B_*<B <B_*+1)  = {\cal Z}_{\rm top}(q,Q)
%{\cal Z}_{\rm top}(q, Q_* Q^{-1})
%$$
%
%where $Q_* = exp(-B_*)$.

\bigskip

%%%%%%%%%

%%%%%%%%%%%%%%%%%%%%%%%%%%%%%%%%%%%%%%%%%%%%%%%%%%%%%%%%%%%%%%%%%%%%%%%%%%%%%%%%
\section{Examples}

In this section we give some examples
of geometries without compact 4-cycles. 
%While we mostly focus on toric Calabi-Yau's our methods are general. 
We first study toric cases, namely resolved conifold and generalized conifolds.
We also give a simple example of a non-compact, non-toric Calabi-Yau as well. In each of these cases, we will use our methods to lay out the chamber structure, identifying walls where BPS states jump, and the BPS partition function in each chamber.
In some of the cases we study, the jumps were studied by other means. We will show that they agree with the M-theory results.

%%%%%%%%%%%%%%%%%%%%%%%%%%%%%%%
\subsection{Resolved conifold}\label{sec.conifold}

The topological string partition function for the resolved conifold
is given by
\beq
 {\cal Z}_{\rm top}(q, Q) = M(q) \prod_{n=1}^{\infty} (1-q^n Q )^n.
\eeq
This means that the only non-vanishing GV
invariants are 
\beq
N_{\beta=\pm 1}^0=1,\quad N_{\beta=0}^0=-2,
\eeq
and that all BPS states in 5 dimensions has no intrinsic spin 
\cite{GV1,GV2}. Our formula \eqref{ZChamber} then implies that BPS states
are counted by
\begin{align}
 {\cal Z}_{\rm BPS} (q, Q) &={\cal Z}_{\rm top}(q, Q){\cal Z}_{\rm top}(q, Q^{-1})|_{\rm chamber} \\
& =  \prod_{(\beta, n): Z(\beta, n)> 0} (1 - q^n Q^\beta)^{n N_\beta^0}.
\label{conifold}
\end{align}
The product is over $\beta =0, \pm 1$ and $n=1,2,...$ such that
$Z(\beta, n)> 0$.

The chamber structure is easy to identify in this case since the K\"ahler 
moduli space is one-dimensional. When
\beq
   R>0 ~ {\rm ~~and~~}~ m-1 < B < m
\eeq
with some $m \geq 1$, the formula \eqref{conifold} gives
\beq
  {\cal Z}_{\rm BPS} (q, Q) = M(q)^2
\prod_{n=1}^\infty(1-q^nQ)^n \prod_{n=m}^\infty
(1-q^n Q^{-1})^n.
\eeq
In particular, the chamber at $m=\infty$ counts the DT invariants
\cite{DT}, 
\beq
  {\cal Z}_{\rm BPS} (q, Q) =  {\cal Z}_{\rm DT} (q, Q) = M(q)^2
\prod_{n=1}^\infty(1-q^nQ)^n,
\eeq
while the chamber at $m=1$ counts the non-commutative DT invariants
\cite{Szendroi},
\beq
  {\cal Z}_{\rm BPS} (q, Q) =  {\cal Z}_{\rm NCDT} (q, Q) = M(q)^2
\prod_{n=1}^\infty(1-q^nQ)^n \prod_{n=1}^\infty
(1-q^n Q^{-1})^n.
\eeq

On the other hand, when 
\beq
   R<0 ~ {\rm ~~and~~}~ -m-1 < B < -m
\eeq
with $m \geq 1$, we have
\beq
  {\cal Z}_{\rm BPS} (q, Q) = \prod_{n=1}^m(1-q^nQ)^n. 
\eeq
In particular, the chamber at $m=\infty$ counts the 
Pandharipande-Thomas invariants \cite{PT},
\beq
  {\cal Z}_{\rm BPS} (q, Q) =  
{\cal Z}_{\rm PT} (q, Q) = \prod_{n=1}^\infty(1-q^nQ)^n. 
\eeq

These agree with the results in 
\cite{Szendroi,JM,CJ,NN} in all chambers.

\subsection{Toric CY without compact 4-cycles}\label{sec.gc}

We can also test our formula \eqref{ZChamber} for a more general
toric CY without compact 4-cycles. 
A toric CY is characterized by a convex polygon on a square 
lattice, and the absence of compact 4-cycles means that there is no
internal lattice point in the polygon. By $SL(2, \mathbb{Z})$ transformations 
of the lattice, one can move one of the edges of the polygon along the
positive $x$-axis, and one of the vertices to $(x,y)$ with 
$-y < x \leq 0$. If we require that there is no internal lattice point,
there are essentially two possibilities: $(x,y) = (0,1)$ and $(0,2)$.
In the former case, the polygon is a trapezoid of height 1, and the
corresponding CY is %a resolution of the $A_{N-1}$ singularity
the so-called generalized conifold, which has $N-1$ $\bP^1$'s 
where $N$ is the area of the trapezoid. We will describe the 
resolved geometry in 
more detail below. In the latter case, 
we have a isosceles right triangle with two legs of length 2,
which corresponds to $\mathbb{C}^3/\mathbb{Z}_2\times \mathbb{Z}_2$.

For the generalized conifold, the topological string partition function 
has been computed in \cite{IK} using the topological vertex \cite{AKMV}.
The counting of BPS states has been carried out in all chambers in 
\cite{Nagao}. Thus, we will use this case to test our formula \eqref{ZChamber}. 
For $\mathbb{C}^3/\mathbb{Z}_2\times \mathbb{Z}_2$, the counting
in the non-commutative DT chamber has been done in 
\cite{Young2}. 

Homology 2-cycles of the generalized conifold %resolved $A_{N-1}$ singularity 
correspond to the simple roots  $\alpha_1, \cdots, \alpha_{N-1}$
of the $A_{N-1}$ algebra. To identify them in the toric diagram, 
we divide the trapezoid into $N$ triangles of area 1 and label
the internal lines dividing the triangles as $i = 1, \cdots, N-1$. 
Each line $i$ corresponds to the blowing up $P^1$ at $\alpha_i$. 
We will denote the D2 charge by
\beq
   \beta = \sum_i n_i \alpha_i.
\eeq
In general, there are several ways to 
divide the trapezoid, and they correspond to different crepant resolutions 
of the singularity. If the two triangles across the line $i$ form 
a rhombus, we have a resolution by $\mathcal{O}(-1,-1)$. On the other hand,
if the two triangles form a triangle of area 2, the resolution is by 
$\mathcal{O}(-2,0)$. Both the topological string partition function 
and the BPS counting depend on the choice of the
resolution. 

The topological string partition function for this geometry is given
by 
\beq
 {\cal Z}_{\rm top} (q, Q)
= M(q)^{N/2} \prod_{n=1}^\infty \prod_{i\leq j} 
(1-q^n Q^i Q^{i+1} \cdots Q^j)^{n N_{ij}}, \label{generaltop}
\eeq
where
\beq
  N_{ij} = (-1)^{1 + n_{ij}}, \label{positiveroot}
\eeq
$n_{ij}= \# \{ k \in I| i \leq k \leq j\}$, and
$I$ is the set of internal lines of the toric diagram 
corresponding to the resolution by
$\mathcal{O}(-1,-1)$. Thus, the only non-vanishing
GV invariants are 
\beq
 N_\beta^{m=0} = (-1)^{1 + \sum_{i\in I} n_i}, \label{root}
\eeq
for a root vector $\beta = \sum_i n_i \alpha_i$
of $A_{N-1}$, and
\beq
N_{\beta=0}^{m=0} = -N.
\eeq
Note that, when $\beta$ is a positive root $\beta_{ij}
= \beta_i + \cdots + \beta_j$, \eqref{root} reduces to 
\eqref{positiveroot}. All BPS states in 5 dimensions carry 
no intrinsic spin. 

The centeral charge $Z(\beta, n)$ is given by
\beq
   Z(\beta, n) = R^{-1}\left(n + \sum_i n_i B_i\right), 
\eeq
where $B_i$ is the B-field evaluated on $\alpha_i$. 
The formula \eqref{ZChamber} predists that BPS
states in the chamber characterized by $B_i$'s are
counted as
\beq
  {\cal Z}_{\rm BPS}(q, Q) = 
  M(q)^N\prod_{(\beta,n): Z(\beta, n)>0} (1-q^n Q^\beta)^{nN_\beta^0}.
\eeq
Here the product is over all roots $\beta$ of $A_{N-1}$ and $n=1,2,...$
such that $Z(\beta, n) > 0$. This agrees with the result in \cite{Nagao}.

\subsection{A Non-toric Example}
\begin{figure}[htbp]
\centering{\includegraphics[scale=0.7]{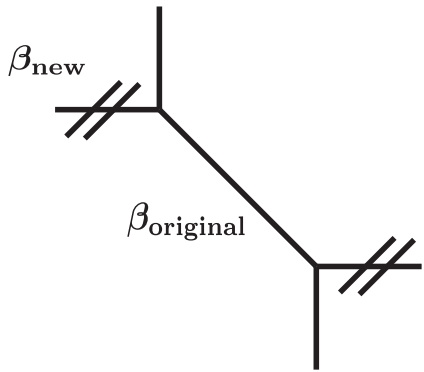}}
\caption{The non-toric CY which arises by identifying two external legs of the $(p,q)$-web of the resolved conifold.}
\label{nontoric}
\end{figure}

Our discussion in sections 2 and 3 are not limited to toric CYs, and applies to any CY without compact 4-cycles. In order to illustrate this point in a concrete setting, let us describe the geometry shown in Figure \ref{nontoric}.
This geometry arises by identifying two of the four external legs of the $(p,q)$-web of the resolved conifold. This is one of the simplest the non-toric geometries studied in \cite{HIV}, and 
it is straightforward to repeat the following analysis to other non-toric geometries discussed in \cite{HIV}.

In addition to the $\bP^1$ of the resolved conifold, the geometry of Figure \ref{nontoric} has another compact $\bP^1$ which arises from identification. Let us denote their homology classes by $\beta_{\rm original}$ and $\beta_{\rm new}$, respectively. As a basis of the homology class, we choose $\beta_1=\beta_{\rm original}$ and $\beta_2=\beta_{\rm original}+\beta_{\rm new}$.

The topological string partition function is given by \cite{HIV}
\beq
\scZ_{\rm top}(q,Q_1,Q_2)=M(q) \left( \prod_{n=1}^{\infty} (1-Q_1 q^n)^n \right) 
\prod_{k,n=1}^{\infty} \left( \frac{(1-q^n Q_1 Q_2^k ) (1-q^n Q_1^{-1} Q_2^k)}{(1-q^{n-1} Q_2^k) (1-q^{n+1} Q_2^k)}  \right)^n
\eeq
where $Q_1$ and $Q_2$ are the variables corresponding to $\beta_1$ and $\beta_2$.
The GV invariants are therefore given by
\beq
N^0_{\beta=0}=-2,\quad  N_{\pm \beta_1}^0=1, \quad N^0_{\pm \beta_1+k\beta_2}=1, \quad
N^{\pm 1}_{k\beta_2}=-1 \quad (k\in \bZ \backslash \{ 0 \}). 
\eeq
Notice that genus 1 GV invariants are nonvanishing in this non-toric example.

Again, the general formula gives (notice that $m\ne 0$ in this case)
\begin{align}
 {\cal Z}_{\rm BPS} (q, Q) &={\cal Z}_{\rm top}(q, Q){\cal Z}_{\rm top}(q, Q^{-1})|_{\rm chamber} \nonumber \\
& =  \prod_{(\beta, l,m): Z(\beta, n=m+l)> 0} (1 - q^{l+m} Q^\beta)^{l N_\beta^m}.
\end{align}
The formula for the central charge is
$$
Z\left(\beta=\sum_{i=1,2} n_i \beta_i,n\right)=R^{-1}\left(\sum_i n_i B_i +n \right),
$$
and the equation 
$$
Z(\beta,n=l+m)=0,
$$ with corresponding GV invariants nonvanishing, determines the position of walls of marginal stability. This is a new result which has not been discussed in the literature to the best of our knowledge.

\section{Relation to the Denef-Moore formula}
In this final section, we point out that our M-theory viewpoint discussed in this paper allows us to derive, in the present context of D6, D2 and D0 degeneracies, the ``semi-primitive" wall crossing formula of \cite{DM}.
The latter says the following. Suppose a BPS bound state of charge $\gamma$ decays into two fragments.
Since the D6 brane is non-compact and fills the entire CY,
the fragments should have charges
$\gamma_1=(1,0,\beta^{'},n^{'})\in H_6\oplus H_4\oplus H_2\oplus H_0$ and $\gamma_2=(0,0,\beta,n)$
\cite{JM}.
The position of walls is determined by the condition that the central charges align
\beq
{\rm Im}(Z(\gamma_1) \overline{Z(\gamma_2)} )=0, \label{wallposition}
\eeq
where $Z(\gamma_i)$ are the central charges for
the D brane charges $\gamma_i$.
The prediction of \cite{DM} is that across such a ``semi-primitive" wall
the partition function jumps by a factor\footnote{We here consider cases where GV invariants are vanishing except for genus 0, such as the toric examples discussed in sections \ref{sec.conifold} and \ref{sec.gc}.}
$$
(1 -q^n Q^\beta)^{n N({\beta},n)}.
$$
The fact that the same factors enter the topological string partition
begged for an explanation. We have provided it by using STS duality to relate both the topological string and the D6-D2-D0 degeneracy counting to M-theory, where the computations unify. Note however that, while the topological string computes the pieces of the D6-D2-D0 degeneracies, the two partition functions are the same only in one chamber.

%%%%

\section*{Acknowledgments}

We would like to thank G.~Moore, H.~Nakajima, K.~Nagao for stimulating discussions.

M.~A. is supported in part by the UC Berkeley Center for Theoretical 
Physics and the NSF grant PHYS-0457317. 
H.~O. and M.~Y. are supported in part by DOE grant DE-FG03-92-ER40701 and by the
World Premier International Research Center Initiative of MEXT of Japan.
H.~O. is also supported in part by a Grant-in-Aid for Scientific
Research (C) 20540256
of JSPS and by the Kavli Foundation. 
C.~V. is supported in part by NSF grant PHY-0244821. 
M.~Y. is also supported in part
by the JSPS fellowships for Young Scientists and by the
Global COE Program for Physical Sciences Frontier at
the University of Tokyo.

H.~O. thanks the hospitality at the Aspen Center for Physics. C.~V.
and M.~Y. thanks Simons Center for Geometry and Physics at Stony Brook for hospitality.

%%%%%%%%%%%%%%%%%%

\end{document}